\documentclass[]{spie}  

\usepackage{amsmath,amsfonts,amssymb,mathrsfs}
\usepackage{graphicx}
\usepackage[colorlinks=true, allcolors=blue]{hyperref}
\usepackage{astro_bib_macro} 

\title{Coronagraphy for DiRect Imaging of Exoplanets (CIDRE) testbed 1:\\ concept, optical set up, and experimental results of adaptive amplitude apodization}

\author{Lucie Leboulleux\supit{a}, Alexis Carlotti\supit{a}, Stéphane Curaba\supit{a}, Alain Delboulbé\supit{a},  Laurent Jocou\supit{a}, Thibaut Moulin\supit{a}, Laurence Gluck\supit{a}, Marie-Hélène Sztefek\supit{a}
\skiplinehalf
\supit{a} Univ. Grenoble Alpes, CNRS, IPAG, 38000 Grenoble, France
}

\authorinfo{Further author information, send correspondence to Lucie Leboulleux: E-mail: lucie.leboulleux@univ-grenoble-alpes.fr}

\begin{document} 
\maketitle

\begin{abstract}

Oncoming exoplanet spectro-imagers like the Planetary Camera and Spectrograph (PCS) for the Extremely Large Telescope (ELT) will aim for a new class of exoplanets, including Earth-like planets evolving around M dwarfs i.e., closer than $0.1$’’ with contrasts around $10^{-8}$. This goal can be achieved with coronagraphs to modulate the incident wavefront. However classical coronagraphs are not optimal: 1) they impose a planetary photon loss, which is particularly problematic when the instrument includes a high spectral-resolution spectrograph, 2) some aberrations such as the missing segments of the ELT are dynamic and not compatible with a static coronagraph design, 3) the coupling of the exoplanet image with a fiber for spectroscopy only requires the electric field to be controlled on a small target-dependent region of the detector. 

Such instruments would benefit from an adaptive tool to modulate the wavefront in both amplitude and phase. We propose to combine in the pupil plane a deformable mirror (DM) to control the phase and a digital micro-mirror device (DMD) i.e., an array made of $1920 \times 1080$ micro-mirrors able to switch between two positions, to control its amplitude. If the DM is already well-known in the field in particular for adaptive optics applications, the DMD has so far not been fully considered. At IPAG, we are currently assembling a testbed called CIDRE (Coronagraphy for DiRect Imaging of Exoplanets) to develop, test, calibrate, and validate the combination of these two components with a Lyot coronagraph.

Since March 2022, CIDRE is assembled albeit without the Lyot coronagraph yet. The first few months have been dedicated to the calibration of the DMD. Since May 2022, it is operational and used to test dynamic amplitude apodization coronagraphs (so-called Shaped Pupils). This proceeding presents the set up of the CIDRE testbench and the first experimental results on adaptive Shaped Pupils obtained with the DMD.

\end{abstract}


\section{INTRODUCTION}
\label{sec:INTRODUCTION}

\begin{figure}
   \begin{center}
   \begin{tabular}{c}
   \includegraphics[width=12cm]{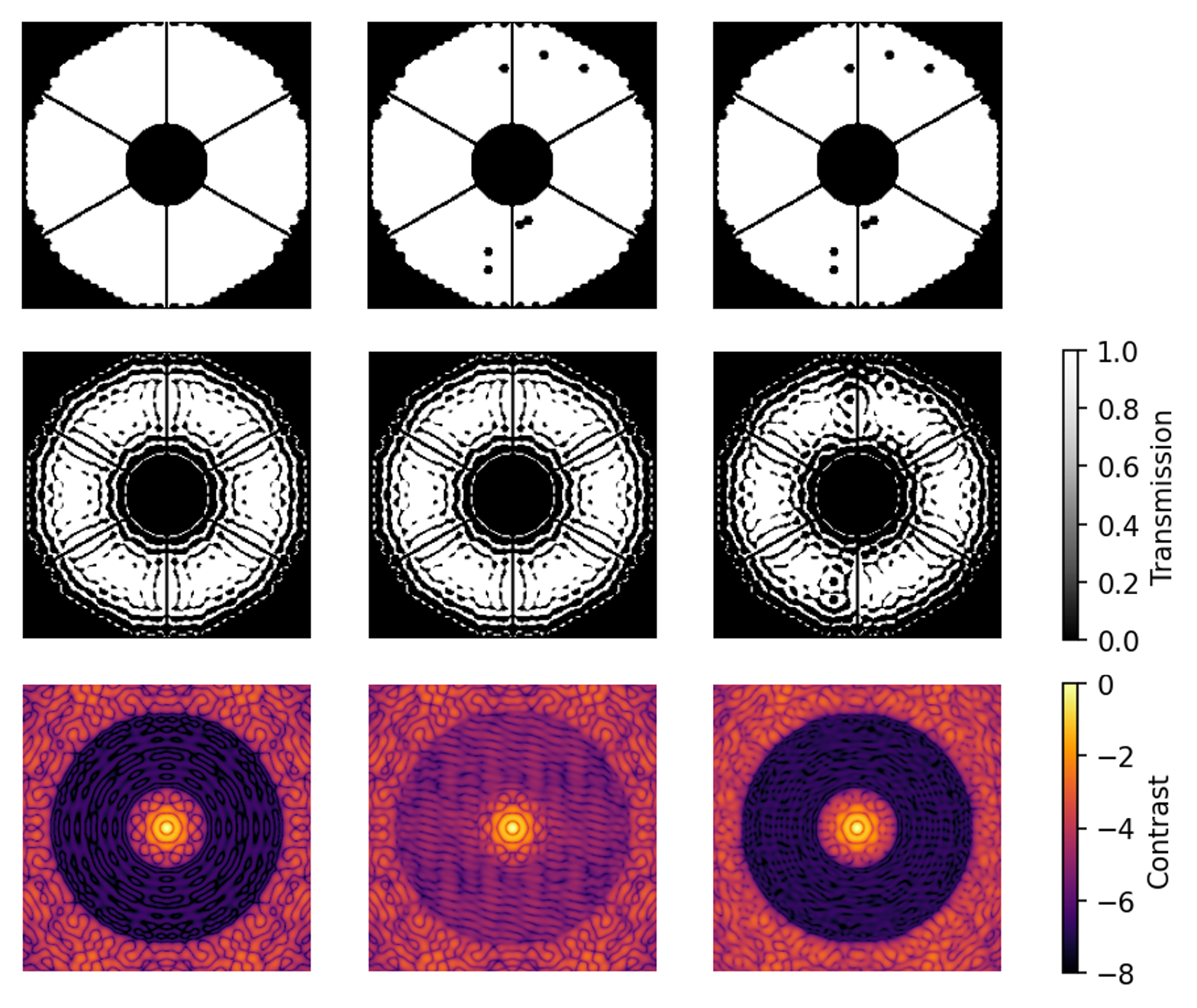}
   \end{tabular}
   \end{center}
   \caption[Fig1] 
   { \label{fig:Fig1} Interest of an adaptive apodization: (top) primary mirror configuration, (center) amplitude apodization (Shaped Pupil), (bottom) contrast map in logarithmic scale with the optical configuration of the first two lines. In the best scenario (left) the ultimate performance is obtained with a contrast of $1.8 \times 10^{-7}$ between $8$ and $20 \lambda/D$ with a planetary throughput of $38\%$. In presence of missing segments (center), as expected on the ELT, this contrast is deteriorated to $9.6 \times 10^{-6}$. The ultimate performance can be recovered despite these missing segments by updating the apodization (left), with a contrast of $1.4 \times 10^{-7}$ and a throughput of $31\%$.}
   \end{figure}

The characterization of young exoplanets with future Extremely Large Telescope (ELT) instruments will rely on coronagraphs accessing small angular separations and combined with high spectral-resolution spectrometers. However, coronagraphs are highly sensitive to aberrations, and missing segments in the primary mirror as expected on the ELT can deteriorate the contrast and so the access to target planets \cite{Carlotti2018a, Mazoyer2018, Mazoyer2018a, Laginja2021, N'Diaye2015, Yaitskova2003, Stahl2020, Leboulleux2018}. As an example, Fig.~\ref{fig:Fig1} illustrates the evolution of a coronagraph performance for an ELT-like telescope without (left) and with (center) missing segments: the contrast in the coronagraphic dark zone is deteriorated by a factor of $50$. Adapting the apodization to the real pupil configuration, including the missing segments, enables to recover the target performance (right).

\begin{figure}
   \begin{center}
   \begin{tabular}{c}
   \includegraphics[width=12cm]{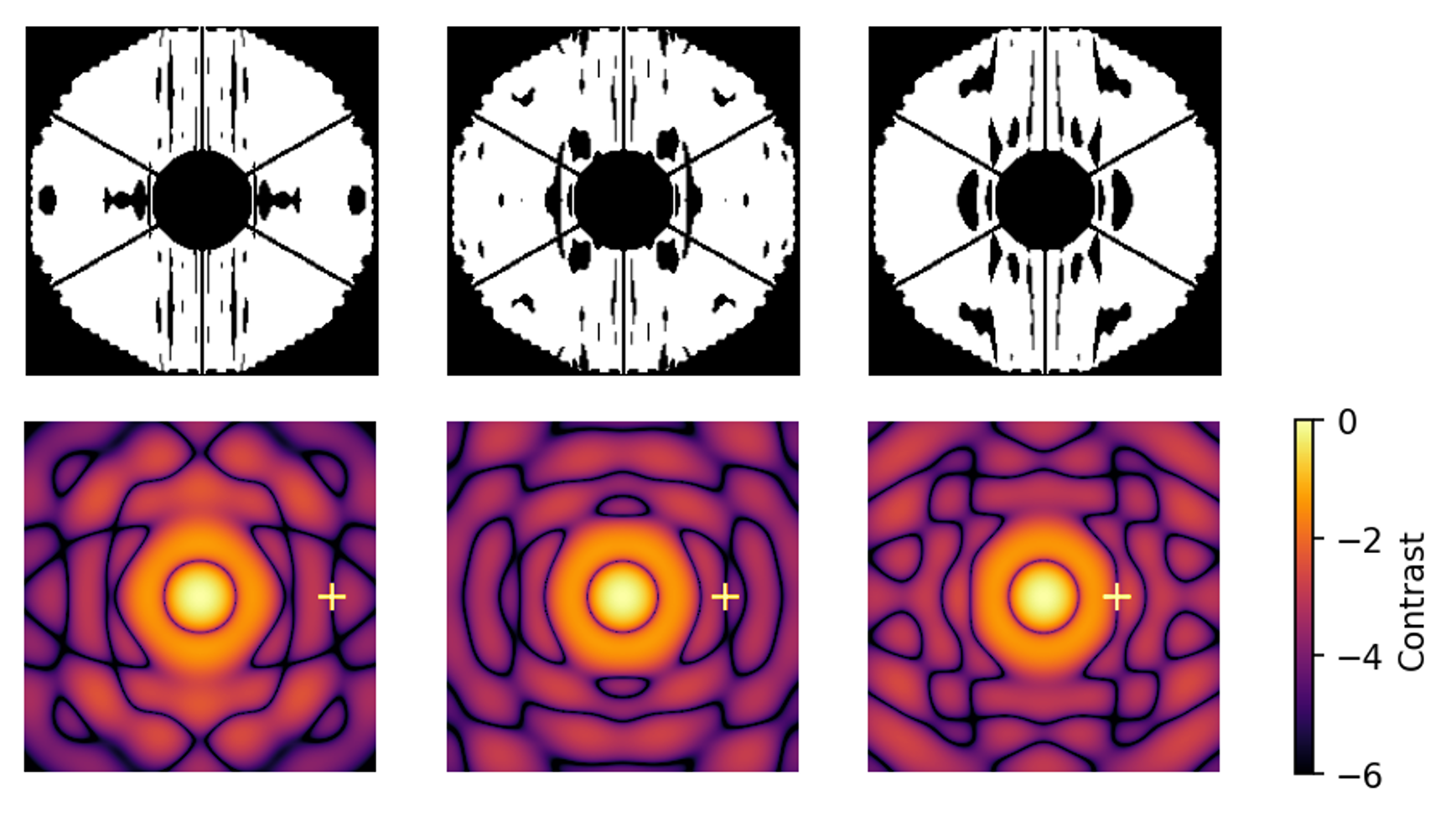}
   \end{tabular}
   \end{center}
   \caption[Fig1bis] 
   { \label{fig:Fig1bis} Interest of an adaptive apodization: (top) amplitude apodization (Shaped Pupil), (bottom) fiber-coupling contrast map in logarithmic scale with the Shaped Pupils of the first line. The planet and single mode fiber positions are indicated with a cross and move closer to the star: (left) $4.5 \lambda/D$, (center) $3.5 \lambda/D$, (right) $2.5 \lambda/D$.}
   \end{figure}

In addition, an instrument currently has to include different coronagraphs to access different angular separations or contrasts: for instance, the Spectro-Polarimetric High-contrast Exoplanet Research (SPHERE) instrument at the Very Large Telescope (VLT) has two apodizers and three Focal Plane Masks (FPM) \cite{Boccaletti2008, Carbillet2011, Guerri2011, Beuzit2019}, or the High-Contrast Module (HCM) of the High Angular Resolution Monolithic Optical and Near-infrared Integral field spectrograph (HARMONI) at the ELT will have two amplitude apodizers \cite{Carlotti2018a}. Similarly, the combination of the coronagraph with a single mode fiber for high-resolution spectroscopy does not require a large coronagraphic dark zone but just a control at the fiber location, which evolves with the position of the target planet. Fig.~\ref{fig:Fig1bis} illustrates this effect for an ELT-like configuration, with a planet getting closer to its host star at angular separations from $4.5 \lambda/D$ to $2.5 \lambda/D$, and the apodization enabling to couple its position with a single-mode fiber.

The coronagraphic system should then be compatible to multiple pupil configurations (missing segments, aberrations, etc.) and scientific objectives (evolving dark zone and specifications). In this context, one can use adaptive components. One proposition relies on an apodizer combined with a deformable mirror or on two Spatial Light Modulators (SLM) or deformable mirrors, set both in pupil plane and out of pupil plane to provide phase and amplitude wavefront control \cite{Pueyo2003, Carlotti2013}. For instance, the ACAD-OSM control technique \cite{Pueyo2013, Mazoyer2018, Mazoyer2018a} aims to compensate for the image diffraction patterns due to the pupil discontinuities (spiders, segmentation, missing segments...) with two deformable mirrors.

We propose the use of an alternative technology, the so-called Digital Micro-Mirror Device (DMD) to provide an adaptive amplitude apodization in the coronagraphic instrument directly in the pupil plane. It consists of an array of micron-sized mirrors that can have two orientations to redirect light either to the detector or out of the optical train. Therefore, it can generate binary patterns such as amplitude apodization.

An optical testbed, called CIDRE (Coronagraphy for DiRect Imaging of Exoplanets) has been built at the Institute of Planetology and Astrophysics of Grenoble (IPAG) to test and calibrate this component. It also includes a Deformable Mirror (DM) conjugated with the DMD: for now, this DM is only used to compensate for the static low-order phase aberrations induced by the DMD and optical errors. Later on, it will also generate phase apodization \cite{Carlotti2018} and, combined with the DMD, it will provide complex apodization \cite{Carlotti2013, Fogarty2018}. Planned activities include:

\textbullet~the implementation of Shaped Pupils (SP) with the DMD for large and small dark zones \cite{Vanderbei2003, Kasdin2003, Carlotti2011},

\textbullet~its \hspace{0.01cm} combination \hspace{0.01cm} with \hspace{0.01cm} a \hspace{0.01cm} Lyot \hspace{0.01cm} coronagraph \hspace{0.01cm} to \hspace{0.01cm} validate \hspace{0.01cm} adaptive \hspace{0.01cm} Apodized \hspace{0.01cm} Pupil \hspace{0.01cm} Lyot \hspace{0.01cm} Coronagraphs \\ (APLC) \cite{Soummer2009, N'Diaye2015, Zimmerman2016},

\textbullet~the test of complex apodization with both the DMD and the DM.

This proceeding introduces the optical set up of the CIDRE testbed and the DMD component (section \ref{sec:Optical design and alignment}) and presents the first experimental tests of adaptive amplitude apodization (SP) with the DMD (section \ref{sec:Validation and first results}).

\section{Optical assembly}
\label{sec:Optical design and alignment}

\subsection{Optical design}

The CIDRE testbed aims to test and validate adaptive hybrid coronagraphs made of:

\textbullet~an adaptive complex apodizer, combining two adaptive components: the DMD for amplitude apodization and the DM for phase apodization, both in pupil planes. The DM is also used to correct for optical wavefront errors, in particular introduced by the DMD.

\textbullet~a set of FPMs,

\textbullet~a set of Lyot stops in a pupil plane.

\begin{figure}[h]
   \begin{center}
   \begin{tabular}{c}
   \includegraphics[width=14cm]{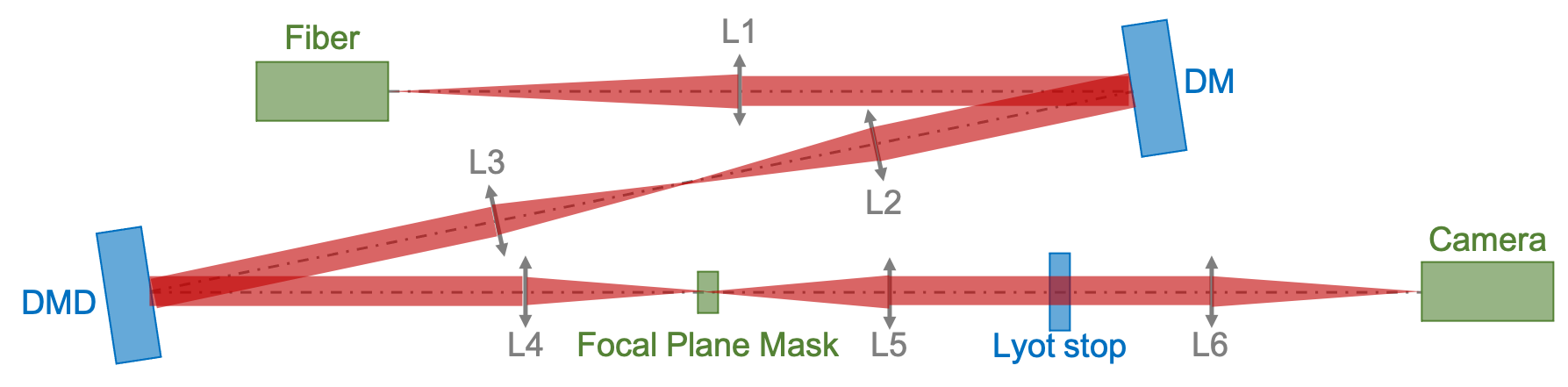}
   \end{tabular}
   \end{center}
   \caption[Fig2] 
   { \label{fig:Fig2} 
CIDRE testbed full design: the DM, DMD, and Lyot stop are set in conjugated pupil planes (blue), the fiber and FPM are set in focal planes (green). The camera can be installed in a focal plane like in this scheme or at the Lyot stop location to image to DMD patterns. The FPMs and Lyot stops are not installed yet.}
   \end{figure}

Its final layout is schemed in Fig.~\ref{fig:Fig2} and includes:

\textbullet~in conjugated focal planes: a laser source, the set of FPMs, and a camera,

\textbullet~in conjugated pupil planes: the DM, the DMD, and the set of Lyot stops. The camera can also be moved to the Lyot stop position to image the pupil,

\textbullet~six lenses in infinity-focus or focus-infinity configurations.

The laser emits a monochromatic light at $635$ nm and all lenses are $1$''-diameter achromatic doublets optimized for visible light. Their specifications (wavelength and diameters) can be found in Table~\ref{tab:FocalLengths} and are constrained by 1) the optical magnification needed so the DM surface (beam size) maps the DM and camera ones, when the camera is set at the Lyot stop position, 2) the resolution of the camera (we obtain $3.9$ pixels per $\lambda/D$), and 3) the size of the optical table.

\begin{table}[!h]
    \caption{Specifications of the lenses.}
    \label{tab:FocalLengths}
    \centering
    \begin{tabular}{l||c|c|c|c|c|c}
        Lens & L1 & L2 & L3 & L4 & L5 & L6 \\
        \hline \hline
        Focal length (mm) & $400$ & $250$ & $200$ & $200$ & $200$ & $250$ \\
    \end{tabular}
\end{table}

\subsection{Component description}

\hspace{0.5cm}\textbullet~\textbf{Deformable Mirror}

\begin{figure}
   \begin{center}
   \begin{tabular}{ccc}
   \includegraphics[height = 6cm]{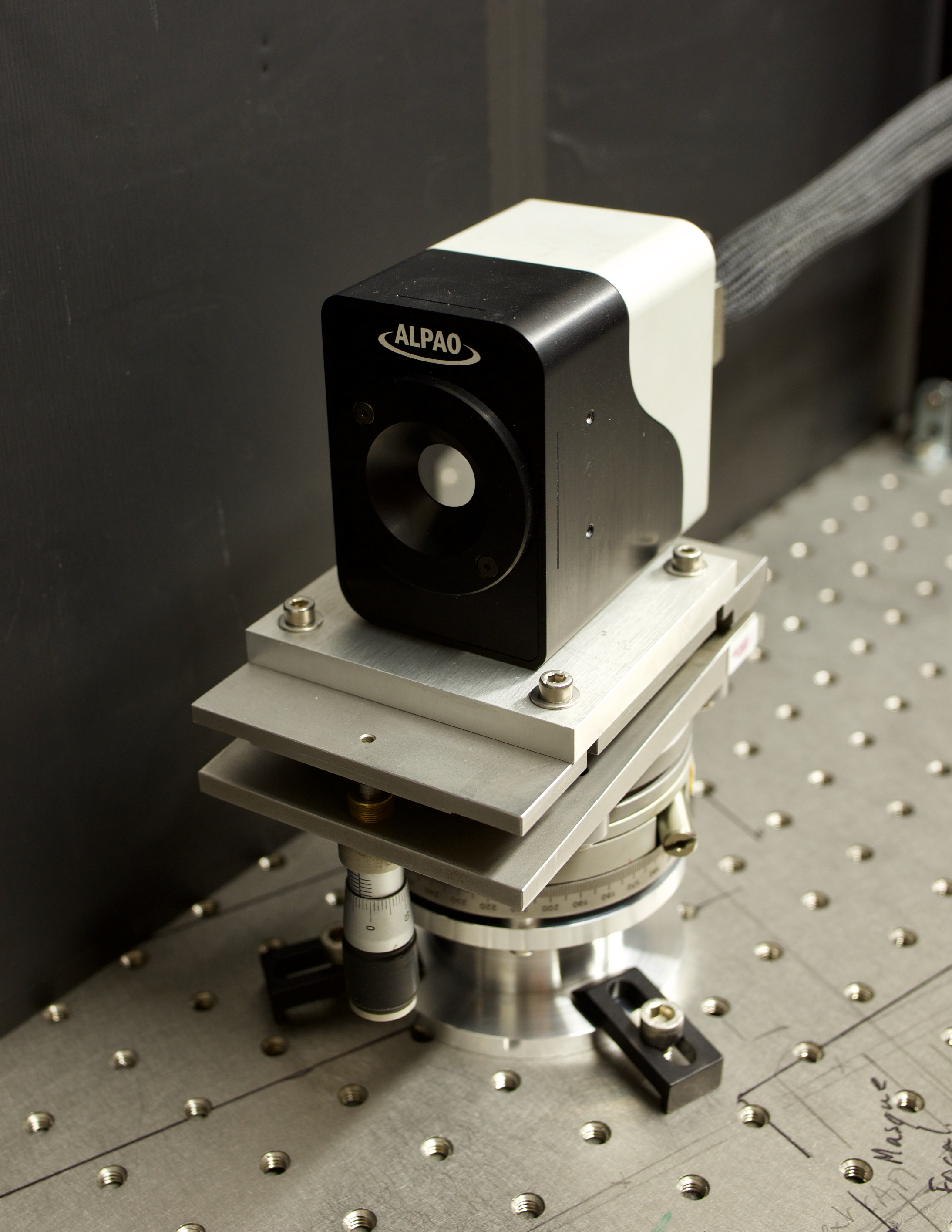} &
   \includegraphics[height = 6cm]{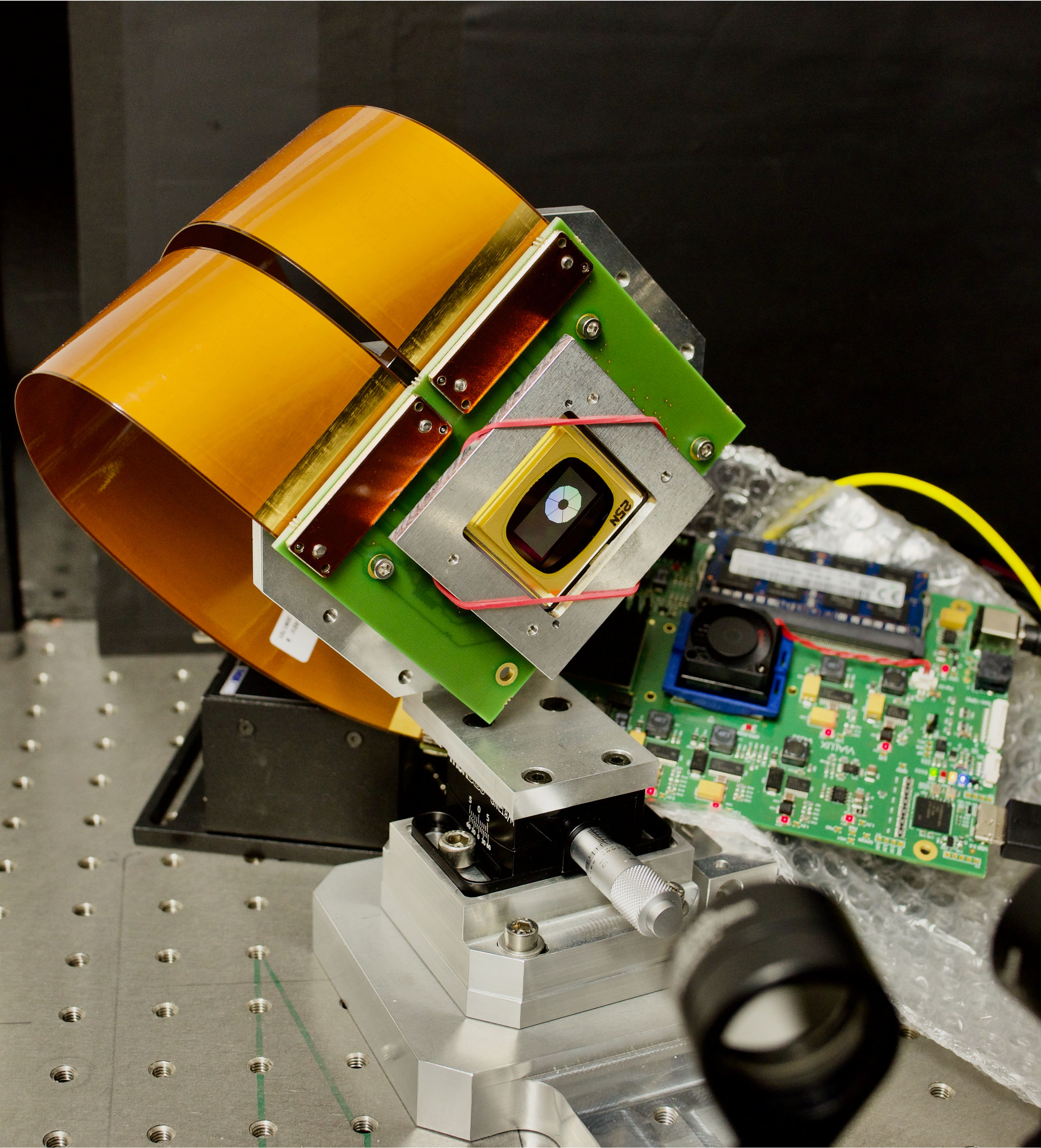} &
   \includegraphics[height = 6cm]{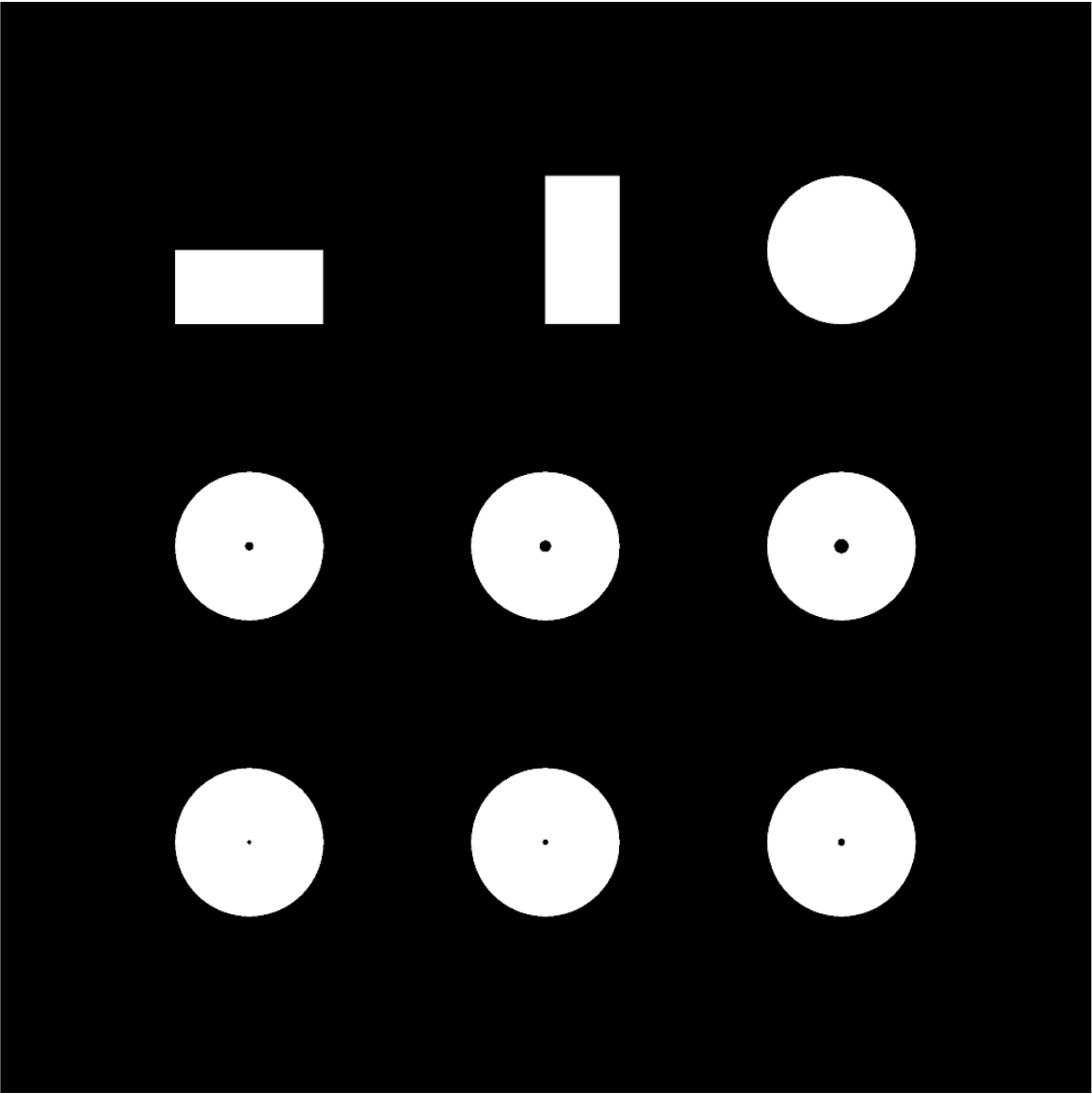}
   \end{tabular}
   \end{center}
   \caption[Fig3] 
   { \label{fig:Fig3} 
CIDRE optical components: (right) ALPAO DM with $97$ actuators, (center) DLP/Texas Instrument DMD with $1920 \times 1080$ micro mirrors, here oriented to map an ELT-like pupil, (left) scheme of the set of FPMs to be ordered to Optimask.}
\end{figure}

The DM set up on CIDRE is the DM97-15 model issued by ALPAO (see also Fig.~\ref{fig:Fig3} (left)). It counts 97 actuators, 11 across the pupil diameter, with a 1.5 mm pitch. The mirror is 13.5 mm large for 7 nm rms of active best flat (specifications provided by ALPAO).

\textbullet~\textbf{Digital Micro-mirror Device}

We chose the Vialux V-9501 model provided by Digital Light Processing (DLP) of Texas Instruments (see also Fig.~\ref{fig:Fig3} (center)), combined with a V4395 controller board.

Its mirror surface is $20.7 \times 11.7$ mm$^2$ and is made of $1920 \times 1080$ micro mirrors, each of them being $10.8$ $\mu$m large. These mirrors can have two orientations by flipping themselves by $\pm 12^{\circ}$ along their diagonal. The DMD fill factor, when all mirrors are oriented towards to optical axis, is $92 \%$, and the mirrors are made of aluminium with a $89 \%$ reflectivity in the visible light \cite{Aswendt2019}.

This micro mirrors tilt generates a periodic structure on the whole mirror surface and therefore a diffraction grating effect following the equation $\sin(\theta_{\pm 1}) = \sin(\theta_0) \pm \lambda/a$ with $a$ at $10.8\sqrt{2}$ $\mu$m. We obtain that the diffraction grating first order is at least at $2.4^{\circ}$ from the optical axis. All non-null orders are stopped by an external radius on the FPMs.

In addition, the DMD surface includes low-order aberrations that have to be compensated with the DM conjugated to the DMD.


\textbullet~\textbf{Focal Plane Masks and Lyot stops}

The FPMs are all printed on the same support and offer different configurations (see Fig.~\ref{fig:Fig3} (right)): horizontal and vertical knife-edge masks and 6 circular masks with different radii, from $1.5$ to $5$ $\lambda/D$. They also have an external mask, with a radius small enough to stop the non-null orders of the DMD-induced dispersion. Similarly, different options of circular Lyot stops will be available, with Lyot ratios from $83$ to $95\%$.

Both FPMs and Loyt stops will be provided by Optimask. For now, the testbed does not include them and only apodization-only coronagraphs can be tested.

\subsection{Testbed assembly}

\begin{figure}
   \begin{center}
   \begin{tabular}{c}
   \includegraphics[width = 16.5cm]{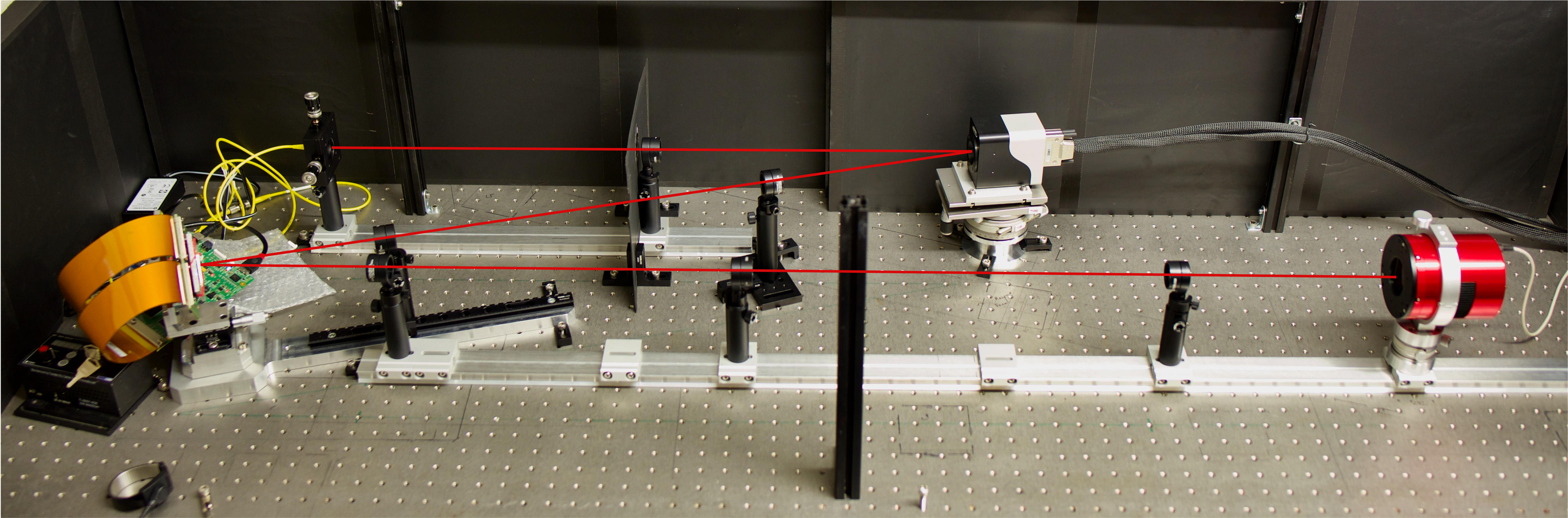}
   \end{tabular}
   \end{center}
   \caption[Fig4] 
   { \label{fig:Fig4} 
Current status of the testbed: all optics (source, DM, DMD, camera, and lenses) are set and aligned, except for the FPMs and Lyot stops. The optical path is indicated as a red line.}
\end{figure}

From February to March 2022, all components have been aligned except for the FPMs and the Lyot stops. Behind the DM, the alignment of all focus-infinity lenses, of the DM and of the DMD is done with a Shack-Hartman wavefront sensor, while all infinity-focus lenses are set with a pinhole and a flat mirror pressed behind them, the reflected light being redirected to the fiber entrance. Additional wavefront flattening is done using the DM, in particular since the DMD surface includes low-order aberrations. At the end of the alignment, the Shack-Hartman wavefront sensor set after the DMD measures less than $0.01$ $\mu$m of any Zernike coefficient between Z5 (astigmatism) and Z14. The residual aberrations being dynamic, a finer alignment does not seem doable in presence of the DMD without DM control in close loop. The current optical assembly is shown in Fig.~\ref{fig:Fig4}.


\section{Validation and first results}
\label{sec:Validation and first results}

\subsection{Full pupil test}
\label{sec:Full pupil test}

The DMD is made of $1920 \times 1080$ micro mirrors and presents an angle compared to the light beam. Any shape to be mapped on the DMD is then centered on the beam impact on the DMD surface, re-shaped to compensate for the reflection angle, and sampled to fit the DMD resolution.

\begin{figure}
   \begin{center}
   \begin{tabular}{c}
   \includegraphics[width = 12cm]{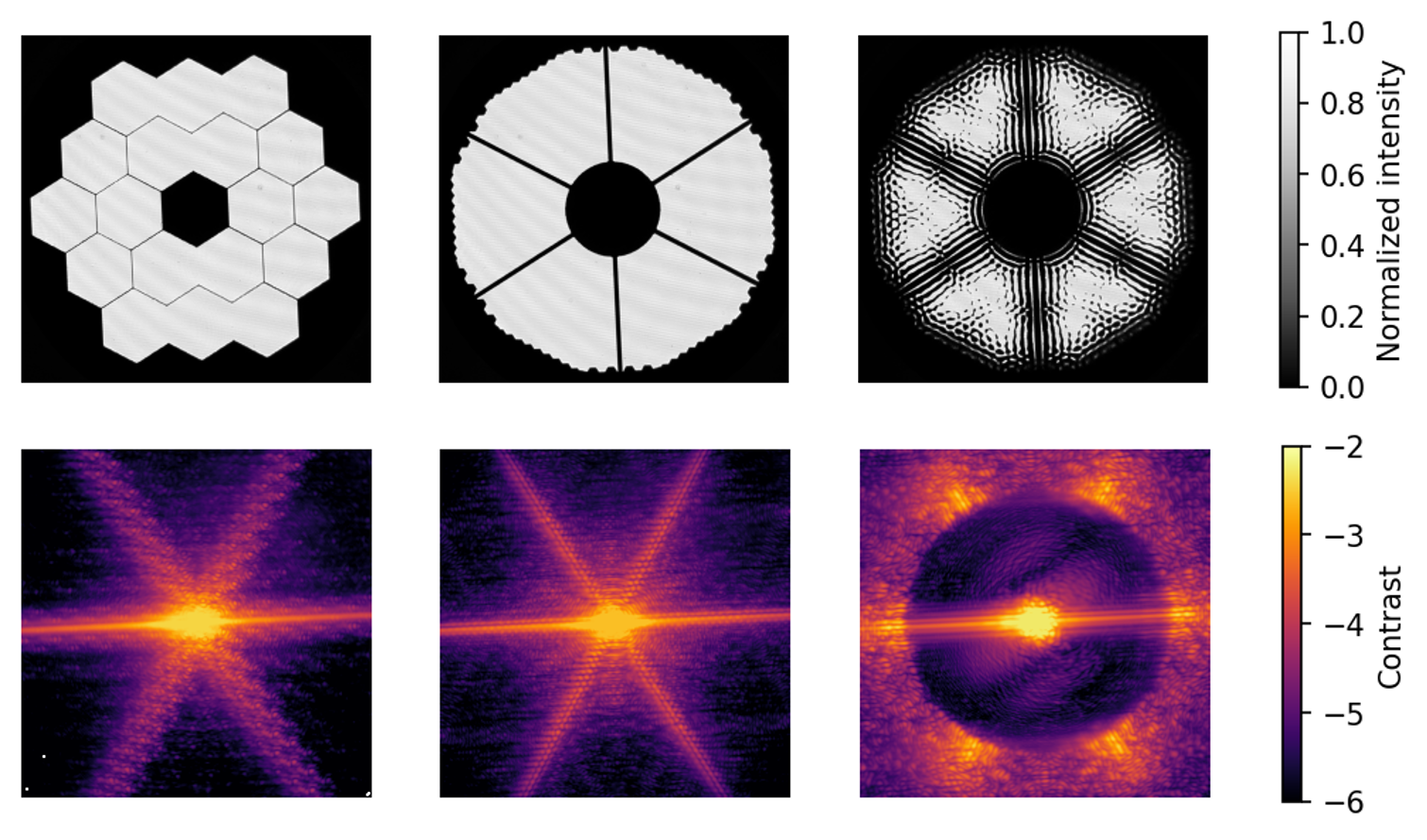}
   \end{tabular}
   \end{center}
   \caption[Fig6] 
   { \label{fig:Fig6} 
First experimental images on CIDRE: (top) images taken with the camera in the pupil plane, conjugated with the DMD surface on which a pupil is mapped, (bottom) associated contrast maps in logarithmic scale. (left) Case of a JWST-like pupil, made of 18 hexagonal mirrors and a hexagonal central obstruction, (center) case of the ELT pupil, (right) case the HSP2 shaped pupil designed for the HCM of ELT/HARMONI designed for a contrast of $10^{-6}$ between $7$ and $40 \lambda/D$.}
\end{figure}

Fig.~\ref{fig:Fig6} presents the first pupils applied on the DMD: a James Webb Space Telescope (JWST)-like pupil, made of 18 hexagonal segments and of a hexagonal central obstruction, the ELT pupil, and the HSP2 shaped pupil designed for the HCM of ELT/HARMONI and aiming for a maximum contrast of $10^{-6}$ between $7$ and $40 \lambda/D$ \cite{Carlotti2018}. 

In the figure, the first line images are taken with the camera set at the Lyot stop plane, so conjugated with the DMD surface and showing what is mapped on its surface. Each image here derives from the average of 20 5-second exposure images, subtracted by 20 equivalent dark images (the laser being off) and divided by flat images (all DMD micro-mirrors reflect the light to the camera).

The second line images are taken with the camera set at the first focal plane behind the DMD (FPM location). Each image here is the average of 100 0.25-second images subtracted by 100 equivalent dark images, and normalized by a non-saturated image computed the same way but with a 650 microseconds exposure time. On the JWST and ELT focal plane images, one can recognize the diffraction peaks due to the segmentation (JWST case) and to the secondary mirror support spiders (ELT case). On the HSP2 focal plane image, the apodization compensates for these diffraction spikes and digs a dark zone. Due to a strong horizontal light leakage, this dark zone is not fully circular but in the two dark zone lobes not impacted by the laser leakage, the mean contrast is $6.3 \times 10^{-6}$.

The light leakage phenomenon, present in all focal plane images and locally deteriorating the contrast, is due to the laser not being completely monochromatic and the DMD acting like a diffraction grating and dispersing light along the horizontal axis. In the future, a new laser should be installed, and further studies should include a co-planar DMD, compatible with broadband light.

\subsection{Adaptive amplitude apodization for missing segments}

One of the main interests of adaptive apodization is its ability to compensate pupil discontinuities such as missing segments, whose configuration will evolve every day on coming large telescopes like the ELT or the TMT (see also Fig.~\ref{fig:Fig1}).

\begin{figure}
   \begin{center}
   \begin{tabular}{c}
   \includegraphics[width = 12cm]{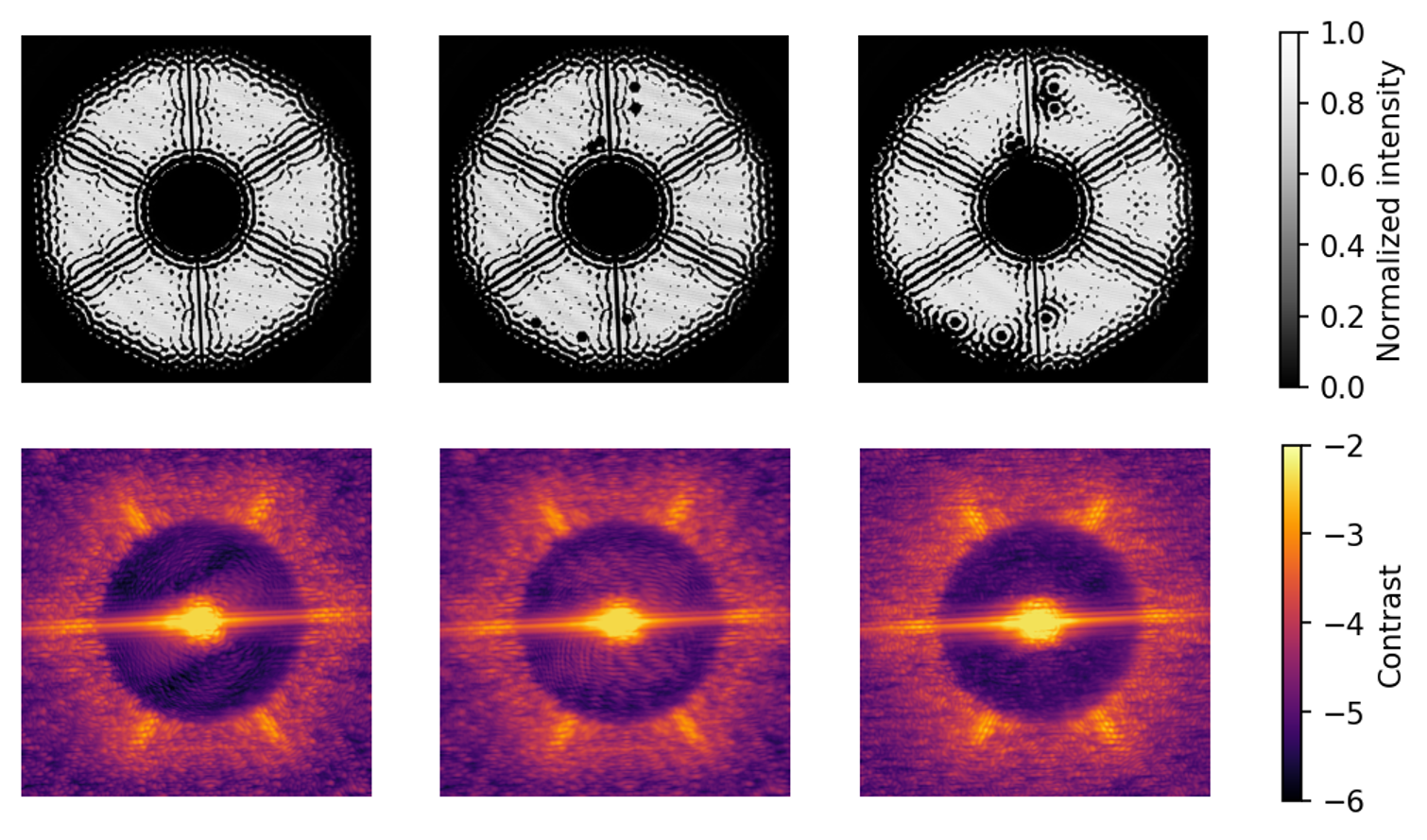}
   \end{tabular}
   \end{center}
   \caption[Fig7] 
   { \label{fig:Fig7} 
Adaptive pupil and apodization scenario on the DMD: (top) images taken with the camera in the pupil plane, conjugated with the DMD surface on which the shaped pupil is mapped, (bottom) associated contrast maps in logarithmic scale. (left) Case of a shaped pupil compatible with the ELT configuration, that aims for an average contrast of $10^{-6}$ between $8$ and $30 \lambda/D$, (center) case of the same shaped pupil combined with missing segments, (right) case the shaped pupil updated to recover the original performance despite missing segments.}
\end{figure}

Fig.~\ref{fig:Fig7} illustrates this scenario with an experimental test on the CIDRE testbed. All images of the figure have the same processing than the ones of Fig.~\ref{fig:Fig6} (section~\ref{sec:Full pupil test}). First, a shaped pupil compatible with the ELT pupil and that targets a maximum contrast of $3.2 \times 10^{-6}$ between $8$ and $30 \lambda/D$ is applied on the DMD. In the experimental set up, an average contrast of $7.5 \times 10^{-6}$ is computed in the two dark zone lobes. This shaped pupil is then combined with missing segments, deriving to a loss of contrast, with an average contrast in the two lobes up to $1.5 \times 10^{-5}$. Eventually a new shaped pupil is designed to be compatible with the ELT pupil including the missing segments, and the target performance is recovered with an average contrast of $6.8 \times 10^{-6}$ in the two dark zone lobes. 

In the focal-plane images, low- to mid-order aberrations remain: they deteriorate both the intensity in the dark zone and the Strehl ratio of the central core and impact the contrast normalization. The next step in this experimental validation should include a fast and efficient wavefront correction procedure. 

\section{Conclusions and perspectives}
\label{sec:Conclusions}

Adaptive coronagraphs would be of high interest on coming large telescopes, with various applications from the high-spectral resolution spectroscopy of a moving target to the adaptation to an optical configuration that evolves with time. To do so, we have identified the DMD technology that can forward light in two different directions with a defined transmission map. 

To experimentally test this technology, we have designed and set up a testbed that is entering its scientific exploitation phase. A few pupils and shaped pupils have been tested, demonstrating the ability of the DMD to shape the intended pattern and to access contrasts down to a few $10^{-6}$. However, several limitations have been identified and will be addressed in the next months:

$\bullet$ the horizontal light leakage is due to the laser being slightly polychromatic and the solution as long as the current DMD is used is to replace it with a fully monochromatic source,
    
$\bullet$ residual low- and -mid order aberrations deteriorate the contrast in the dark zone. A close-loop wavefront control will be implemented in the coming months.
    
More long-term developments are also planned to fully test the concept of adaptative apodization: 

$\bullet$ adding sets of FPMs and of Lyot stops to test adaptive Apodized Pupil Lyot Coronagraphs \cite{N'Diaye2015},

$\bullet$ using the DM as an adaptative phase apodizer to perform both phase and amplitude adaptive wavefront control \cite{Pueyo2003, Fogarty2018},

$\bullet$ identifying an achromatic technology equivalent to the DMD such as a co-planar DMD, to obtain broadband adaptive coronagraphy.

If validated, this concept will be applicable to future spectro-imagers on large telescopes, such as ELT/PCS that will be particularly sensitive to aberrations due to the primary-mirror segmentation while targeting exoplanets with very high contrasts ($10^{-8}-10^{-9}$) \cite{Carlotti2014, Kasper2021}.

\acknowledgments 
This project is funded by the European Research Council (ERC) under the European Union's Horizon 2020 research and innovation programme (grant agreement n°866001 - EXACT).\\
This work has also been partially supported by the LabEx FOCUS ANR-11-LABX-0013.

\bibliography{bib}
\bibliographystyle{spiebib}

\end{document}